\documentclass{aa}
\newcommand{\beq}{\begin{equation}}
\newcommand{\eeq}{\end{equation}}
\def\bea{\begin{eqnarray}}
\def\eea{\end{eqnarray}}

\def\gsim{~\rlap{$>$}{\lower 1.0ex\hbox{$\sim$}}}

\def\simpropto{\lower.2ex\hbox{$\; \buildrel \propto \over \sim \;$}}
\def\ltsim{\lower.5ex\hbox{$\; \buildrel < \over \sim \;$}}
\def\gtsim{\lower.5ex\hbox{$\; \buildrel > \over \sim \;$}}
\def\ltsim{\lower.5ex\hbox{$\; \buildrel < \over \sim \;$}}
\def\gtsim{\lower.5ex\hbox{$\; \buildrel > \over \sim \;$}}

\def\pmb#1{\setbox0=\hbox{#1}%
\kern-.025em\copy0\kern-\wd0
\kern.05em\copy0\kern-\wd0
\kern-.025em\raise.0433em\box0}

\def\simlt{\lower.5ex\hbox{$\; \buildrel < \over \sim \;$}}
\def\simgt{\lower.5ex\hbox{$\; \buildrel > \over \sim \;$}}

\def\fixit#1{}

\def\apjs{ApJ. S}

\def\apjl{ApJL}

\def\aap{A\&A}

\def\apjl{ApJL}

\usepackage{graphicx}
\usepackage{enumitem}
\usepackage{caption}
\usepackage{amssymb}
\usepackage{amsmath}
\usepackage{multirow}
\usepackage{mathrsfs}
\usepackage{bbding}
\usepackage{float}
\usepackage{color}
\usepackage[colorlinks,citecolor=blue]{hyperref}
\newcommand\wx{0.5}
\begin{document}
\title{Evidence of isotropy at large-scale from polarizations of radio sources}
\titlerunning{Evidence of isotropy at large-scale from polarizations of radio sources}
\authorrunning{Tiwari \& Jain}
\author{Prabhakar Tiwari \inst{1} \and Pankaj Jain\inst{2}}
   \institute{ National Astronomical Observatories, CAS, Beijing 100012, China\\
              \email{ptiwari@nao.cas.cn}
             \and
             Department of Physics, Indian Institute of Technology, Kanpur 208016, India\\
             \email{pkjain@iitk.ac.in}
             }
 \abstract
 {We test the isotropy of radio polarization angle orientations with a robust and reliable dual frequency  
 polarimetric survey of active galactic nuclei (AGN). We find that the polarization orientations are 
 consistent with the assumption of isotropy for scales larger than or equal to $\sim 800$ Mpc. 
 This provides further evidence of isotropy at large distance scales and is likely to impose strong constraints 
 on some of the physical mechanisms that may be invoked to explain past observations of alignment of radio
 and optical polarizations at large distance scales.}

   \maketitle
\section{Introduction}
\label{sc:intro}
The universe at large distance scales is assumed to be translational and rotational invariant, avoiding any special point 
or preferred direction in space and thus satisfying the Copernican Principle. 
More generally in modern cosmology we demand the observable universe to be statistically  homogeneous and isotropic 
and this assumption is formally known as  ``Einstein’s Cosmological Principle" \citep{Milne:1933CP,Milne:1935CP}. 
This  is  a fundamental assumption in the standard cosmological framework and therefore must be tested explicitly by observations. Indeed, there are observations 
supporting isotropy, for example the cosmic microwave background (CMB) is uniform in one part in $10^5$ 
\citep{Penzias:1965,COBE_White:1994,WMAP:2013,Planck_iso:2016}, also 
the ultra-high energy cosmic ray (UHECR) events from the Telescope Array (TA) show isotropic distribution 
on the sky \citep{Abu-Zayyad:2012}. Furthermore, the Fermi Gamma-Ray Burst (GRB)  data is also isotopic \citep{Ripa:2017}. Even so there also exist observations indicating large-scale anisotropy 
and most of these remains unexplained till date. In particular the CMB itself shows several anomalies at 
large angular scales \citep{ Oliveira-Costa:2004,Ralston:2004,Schwarz:2004,Schwarz:2016,Aluri:2017,Rath:2018}. The optical polarization from 
quasars shows polarization  angle (PA) alignment  over a very large  scale of Gpc \citep{Hutsemekers:1998}, 
furthermore the radio polarizations also show similar alignment signal at Gpc scales \citep{Hutsemekers:2014}. 

Besides the large scale anisotropies described above, the JVAS/CLASS 8.4-GHz sample \citep{Jackson:2007} of flat-spectrum radio sources (FSRS) 
polarization angles also shows a significant evidence of alignment at distance scale of 
150 Mpc \citep{Tiwari:2013pol}. However the significance of these polarization angle
alignment effects is not very large and requires confirmation with larger data samples. 
With same JVAS/CLASS data \cite{Joshi:2007} report isotropy at relatively larger scales.
 \cite{Shurtleff:2014} reports less significant PA alignments in 
two circular regions of 24$^{\circ}$ radius on the sky. \cite{Jagannathan:2014} also report the jet angle 
alignment across angular scales of up to 1.8 degrees ($\sim$53 Mpc at redshift one) from  
ELAIS N1 Deep radio survey. 
However, after so many different observations and PA alignment studies the situation is 
not  very clear. These anisotropies if real and not some instrumental or observational artifacts
potentially disrupt the modern standard cosmology and thus it is important to review and further investigate 
these claimed anisotropy signals with new and refined observations.  

There exist many theoretical models which aim to explain the observed 
 large scale optical alignment \citep{Hutsemekers:2014}. 
These include mixing of electromagnetic radiation with 
hypothetical light pseudoscalar particles in correlated background 
magnetic fields \citep{Jain:2002vx,Agarwal:2009ic,Piotrovich:2009zz,Agarwal:2012}, vector perturbations \citep{Morales:2007rd}, large scale magnetic field
coupled to dark energy 
\citep{Urban:2009sw}, cosmic strings \citep{Poltis:2010yu,Hackmann:2010ir}, 
anisotropic expansion \citep{Ciarcelluti:2012pc}
and spontaneous violation of isotropy \citep{Chakrabarty:2016}. 
The radio alignment on small distance scales 
\citep{Tiwari:2013pol,Jagannathan:2014} may be explained by correlated supercluster magnetic field \citep{Tiwari:2016}. 
As we shall see our analysis is likely to impose significant constraints on these models. 
  
In this work we employ the simultaneous dual frequency 86 GHz and 229 GHz polarimetric survey \citep{Agudo:2014} 
of radio flux and polarization for a large sample of 211 radio-loud  active galactic nuclei (AGN) to test the 
hypothesis that the polarization vectors of AGNs are randomly oriented at large-scales in the sky. These are 
clear and reliable measure of radio polarizations from a dedicated polarimetric survey at two frequencies 
simultaneously and thus an excellent catalog to test isotropy at large-scales. 

The outline of the paper is as follows. We provide the  details about the survey and data in Section \ref{sc:data}.
In Section \ref{sc:stat} we discuss the methods to measure the isotopy and quantify the anisotropy signal. 
We present the analysis and our detailed results in Section \ref{sc:res}. In Section \ref{sc:con} we discuss and 
conclude this work. 

\section{Data sample}
\label{sc:data}
\begin{figure}
	\includegraphics[width=\wx\textwidth]{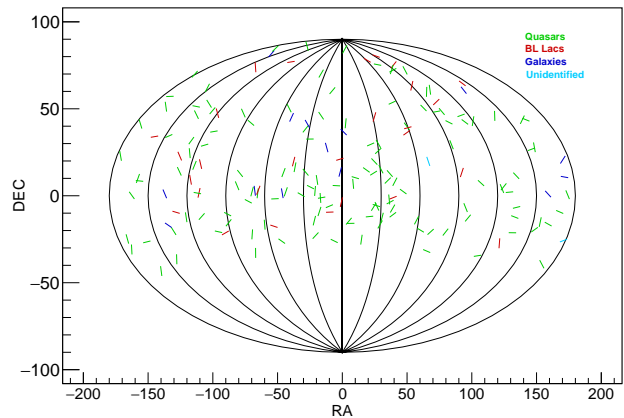} %
	\caption{The sky distribution of sources and their respective polarization angles measured with 
	respect to the local longitude.}
	\label{fig:dist}
\end{figure}

\begin{figure}
	\includegraphics[width=\wx\textwidth]{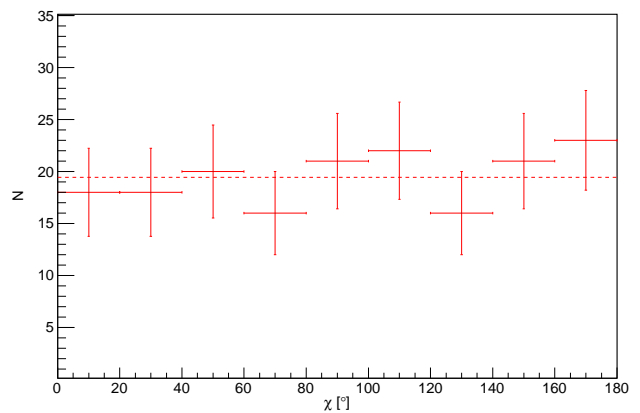}%
	\caption{The polarization angle distribution in sample}
	\label{fig:PAs}
\end{figure}

The dual frequency 86 GHz and 229 GHz catalog we use contains 221 radio-loud AGNs \citep{Agudo:2014}. The observations 
were performed on the IRAM (Institute for Radio Astronomy in the Millimeter Range) 30 meter telescope with the
XPOL polarimeter \citep{Thum:2008}. The sample AGNs are flux limited and  are above 1 Jy (total-flux) at 86 GHz, 
the redshift of 199 AGNs is known and ranges from $z=0.00068$  to 3.408. The mean and median redshift 
of the sample is 0.937 and 0.859, respectively. The sample is dominated by  blazars and contains 
152 quasars, 32 BL Lacs and 21 radio galaxies and 6 unclassified sources. The linear polarization above 
3$\sigma$ median level of $\sim$1\% is detected for 183 sources. In the sample, for 22 sources the linear 
polarization angle is measured at both 86 and 229 GHz, and a good match between the PAs at these frequencies is 
seen (see Fig. 14 in \citealt{Agudo:2014}). The sample is dominantly in the northern sky and covers the entire 
northern hemisphere nearly uniformly. We have shown the sky distribution 
of sources in Fig. \ref{fig:dist}. 
The short millimeter survey is an excellent probes of 
radio loud AGN jets and has several advantages over radio centimeter i.e. $\sim$GHz surveys. 
The millimeter radiation is predominately from the cores of synchrotron-emitting relativist jets and has 
negligible contribution from host AGN and its surroundings. It is also less affected by 
Faraday rotation and depolarization \citep{Zavala:2004,Agudo:2010}. The PAs in the sample are 
uniform between $0$ to $180^{\circ}$, their distribution is shown in Fig. \ref{fig:PAs}. 
The jet position angle with respect to  polarization angle in sample are not preferably 
parallel or perpendicular \citep{Agudo:2014}. In general the survey sample is robust and no 
suspicious systematics or unusual behaviour is observed. Further details 
about the observation and calibration can be found in \cite{Agudo:2014}. 

\section{Measure of anisotropy}
\label{sc:stat}
\begin{figure}
        \includegraphics[width=\wx\textwidth]{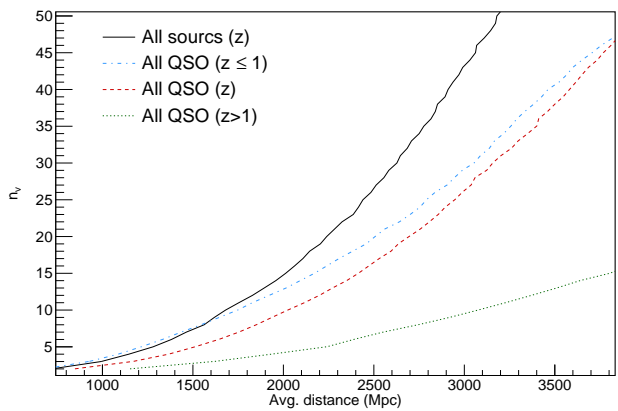}%
        \caption{The average distance corresponding to $n_v$. The distance is computed assuming 
	$\Lambda$CDM and the cosmological parameters are taken from the latest Planck results \citep{Planck:2018}.
	`All sources (z)' sample contains 175 sources, all these have redshifts and good polarization measurements.
	139 out of these 175 sources are quasars, 69 quasar are above redshift one.}
        \label{fig:nv}
\end{figure}

The polarization angles in sample are on the hypothetical  celestial sphere and directional
measurement on this sphere corresponds to a particular coordinate system. To test our hypothesis 
of isotropy, we resort to \cite{Jain:2003sg} coordinate independent statistics procedure 
and compare the polarization vector of a source
after transporting it to the position of a reference source along the geodesic joining the two.
We define the measure of isotropy as following 
\citep{Hutsemekers:1998,Jain:2003sg}. 
We have redshift and angular position i.e. right ascension and declination given for each source. We define the separation between any two given sources in terms of comoving distance. For calculating comoving distances we have assumed $\Lambda$CDM and cosmological parameters from the latest {\it Planck} results \citep{Planck:2018}. Next we consider the $n_v$ nearest neighbours of a source situated at 
site $k$ with its polarization angle $\psi_k$. Let $\psi_i$  be the polarization angle of 
the source at $i^{\rm {th}}$ site within the $n_v$ nearest neighbour set.  A dispersion measure 
of polarization angles relative to $k^{\rm {th}}$ source  $\psi_k$ with its $n_v$ nearest 
neighbour is written as, 
\bea
d_{k} = \frac{1}{n_{v}} \sum_{i=1}^{n_{v}} \cos[2(\psi_{i}+\Delta_{i\rightarrow
k}) - 2{\psi}_{k}],
\label{eq:dispersion}
\eea
where $\Delta_{i\rightarrow k}$ is a correction to angle 
$\psi_{i}$ due to its parallel transport from site $i\rightarrow k$ \citep{Jain:2003sg}. 
The polarization angles span a range of $0$ to $180^\circ$ and to make them behave like 
usual angles and take values over entire range $0$ to $360^\circ$ we multiply the polarization 
angles by two \citep{Ralston:1999}. The function $d_k$ is the average of the cosine of the differences of the polarization 
vector at site $k$ and those of its $n_v$ nearest neighbour set and hence is a measure of the dispersion in angles. It takes on higher values for data with lower dispersions and vice versa.
We take the average of $d_k$ over all source sites and define this 
as a measure of alignment in sample, 
\bea
S_{D} = \frac{1}{N_t} \sum_{k=1}^{N_{t}} d_{k},
\label{eq:statistics}
\eea
where $N_t$ is the total number of sources in the sample.  Similar alternate statistics 
can also be defined as a measure of alignment 
\citep{Bietenholz:1986,Hutsemekers:1998,Jain:2003sg,Tiwari:2013pol,Pelgrims:2014}. 
The nearest neighbour statistics uses number of nearest neighbour, $n_v$, as a proxy of 
distance. This is to fix {\it statistics} at each source location i.e. to have same $n_{v}$ while 
calculating $d_{k}$. The average distance corresponding to $n_v$ for different sub-samples studies in this 
work is given in Fig. \ref{fig:nv}. 

The error in the statistic $S_{D}$ is computed using jackknife method \citep{Tiwari:2016} and the significance is 
computed by comparing the random sample $S_{D}$ with the data $S_{D}$. The jackknife error on $S_{D}$ 
for a given $n_v$ is calculated by re-sampling the data. We eliminate the $i^{{\rm th}}$ source from the
full sample and calculate the correlation statistics  $S_D(i)$. Given full sample statistics $S_D$
the jackknife error $\delta S_D$ in its estimation is  given as, 
\begin{equation}
(\delta S_D) ^2 = \frac{(N_t-1)}{N_t}\sum^{N_t}_{i=1} (S_D(i) -S_D)^2. %
\label{eq:eSD}
\end{equation}

\section{Analysis and results}
\label{sc:res}
\begin{figure}
	\includegraphics[width=\wx\textwidth]{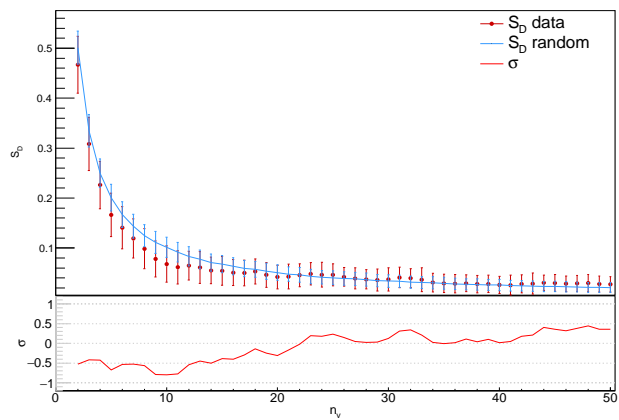}%
	\caption{The statistics $S_D$ of all sources and its significance with respect to random isotropic
	polarization distribution. The error bars on data $S_D$ are jackknife errors and the random 
	polarization angle $S_D$ is also shown with its RMS errors bars drawn from 1000 random samples. The 
	$\sigma$ significance is the difference between data $S_D$ and random $S_D$ as defined in Eq. \ref{eq:sigma}.}
        \label{fig:SD}
\end{figure}

\begin{figure}
        \includegraphics[width=\wx\textwidth]{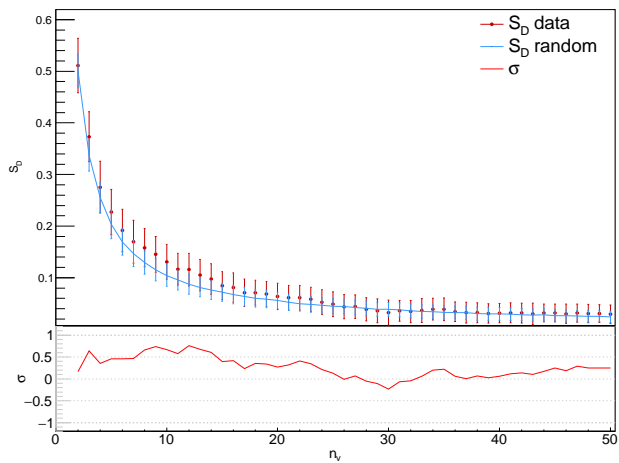}%
        \caption{The statistics $S_D$ of jet angles. Jet position angles are fairly isotropic and agree well 
	within one $\sigma$ with random sample. Other details same as in Fig. \ref{fig:SD}}.
        \label{fig:SD_jet}
\end{figure}

Our full sample contains 211 sources; 175 out of these have both redshift and good polarization 
measurements and so a good sample for our analysis. There are 183 sources with jet angles 
measurements along with redshifts. The $S_D$ measurements and its significance as compared to 
random sample are shown in Fig. \ref{fig:SD}. The $\sigma$ significance i.e. how the sample 
behaves with respect to random isotropic and uniform polarization angles is defined as follow, 
\begin{equation}
	\sigma = \frac{S_D (\rm {data})- S_D(\rm {random})}{\sqrt{(\delta S_D^{\rm data})^2 + (\delta S_D^{\rm random})^2}} %
\label{eq:sigma}
\end{equation}
where $\delta S_D^{\rm data}$ and $\delta S_D^{\rm random}$ are jackknife errors on data $S_D$ and RMS 
errors on random $S_D^{\rm random}$, respectively.  The results with full data in Fig. \ref{fig:SD} shows no 
significant deviation from isotropy and are well within one sigma of the isotropic and uniform
distribution of polarization angles. The jet angles also show good agreement with random jet angle
distribution, the results are shown in  Fig. \ref{fig:SD_jet}. The average distance to first nearest 
neighbour in full  sample is 719 Mpc, 
thus the observation of isotropy in this work applies for distance scales $\ge 719$ Mpc. We are limited by source 
number density to probe smaller scales and so can't explore the polarization angle alignment signal  claimed 
at scales of order 100 Mpc \citep{Tiwari:2013pol}.

\begin{figure}
        \includegraphics[width=\wx\textwidth]{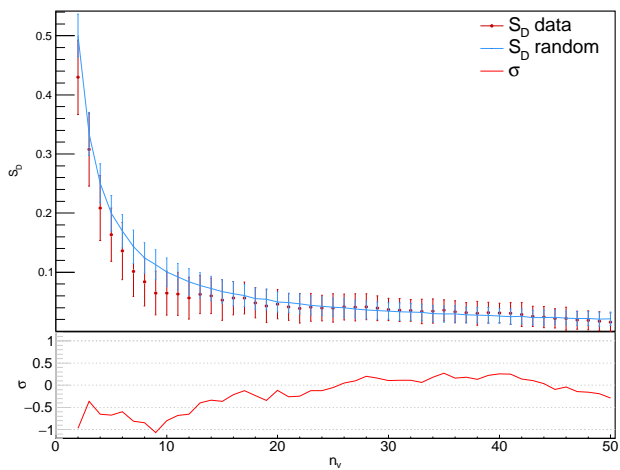}%
        \caption{The statistics $S_D$ exclusive to quasar sample. Other details same as in Fig. \ref{fig:SD}}.
        \label{fig:SD_QSO}
\end{figure}

There have been several observations of quasar polarization alignments at large distance scales 
\citep{Hutsemekers:2014,Pelgrims:2015,Pelgrims:2016} and so we also explore the anisotropy in
quasars only sample. We have in total 139 quasars with redshift and measure 
of  polarization angles in our full sample. These are evenly distributed over northern sky 
(Fig. \ref{fig:dist}) and centered around redshift one (see figure 2 and 3 in \citealt{Agudo:2014}). 
Again, we do not see any alignment in this quasar only sample and the statistics agrees well with random 
distribution. The results are shown in Fig. \ref{fig:SD_QSO}. The least nearest neighbour 
distance in this sample is 849 Mpc and so the quasar polarization angle distribution is isotropic  
at least at this scale and above. 

\begin{figure}
        \includegraphics[width=\wx\textwidth]{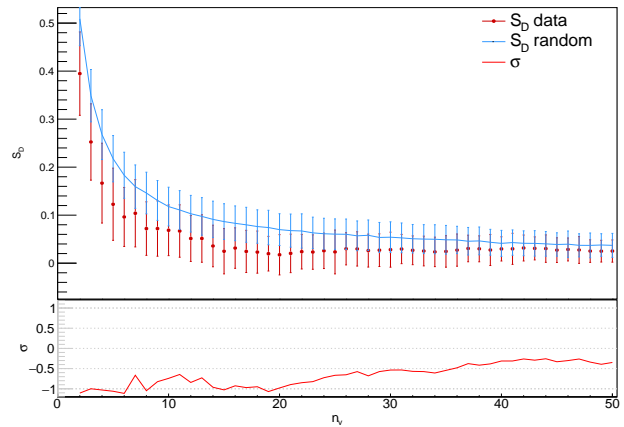}%
        \caption{The quasar sample with redshift greater than one. Other details same as in Fig. \ref{fig:SD}}.
        \label{fig:SD_zg1}
\end{figure}

Next, we test if the alignment signal is redshift dependent, and if it is present with 
high redshift quasars \citep{Pelgrims:2015}. We consider quasars with redshift larger than one and calculate 
$S_D$, this sample contains 67 quasars and for this sample the average distance to first nearest neighbour is 
1153 Mpc. We find that this sample  also agrees well with random, no significant alignment is seen, the results
are shown in Fig. \ref{fig:SD_zg1}. The quasar sample with redshift less than one is also consistent 
with isotropic polarization distribution.


\section{Discussion and Conclusion}
\label{sc:con}
The polarization angle alignment of radio sources and quasars at large distance scales remains puzzling since very long 
\citep{Birch:1982,Kendall:1984,Hutsemekers:1998,Jain:1999, Tiwari:2013pol,Hutsemekers:2014,Pelgrims:2015, Pelgrims:2016}.
Surprisingly, most of these along with cosmic microwave background (CMB) dipole-quadrupole-octopole 
\citep{Oliveira-Costa:2004,Schwarz:2004}, radio galaxy distribution dipole 
\citep{Singal:2011,Gibelyou:2012,Rubart:2013,Tiwari:2014ni,Tiwari:2015np,Tiwari:2016adi,Colin:2017},  and 
polarizations at optical frequencies at cosmological scale indicate a preferred direction pointing roughly towards the
 Virgo cluster \citep{Ralston:2004}. Although there exist some explanations for the large-scale optical 
polarization alignment following axion-photon interaction \citep{Agarwal:2012}, and radio polarization 
alignment at 150 Mpc scale in terms of galaxy supercluster magnetic field \citep{Tiwari:2016}; the alignment 
signal reported in \cite{Hutsemekers:1998,Hutsemekers:2014,Pelgrims:2015, Pelgrims:2016} remains puzzling. 
In this work we explore on large-scale millimeter radio polarization alignment and  report isotropy. 

The observed isotropy is likely to impose significant constrains on the proposed mechanisms for large scale polarization alignment, discussed in 
the Introduction \citep{Jain:2002vx,Agarwal:2009ic,Piotrovich:2009zz,Agarwal:2012,Morales:2007rd,Urban:2009sw,Poltis:2010yu,Hackmann:2010ir,Ciarcelluti:2012pc,Chakrabarty:2016}. The constraints may be more stringent on mechanisms which predict an effect independent of frequency or a larger effect at smaller frequency. This is because such models will predict an alignment of same or higher strength in comparison to that seen at optical frequencies \citep{Hutsemekers:1998}. In contrast the constraints may not be very stringent on a mechanism, such as pseudoscalar-photon mixing \citep{Jain:2002vx}, which increases with frequency. Many of these mechanisms rely on the presence of large scale correlated magnetic field. Hence the observed alignment may also be a probe of the correlations in the large scale magnetic field. However this relationship between alignment and magnetic field is not direct since it also relies on some other effect, such as the mixing of electromagnetic radiation with hypothetical pseudoscalar particles \citep{Andriamonje:2007,Tiwari:2012ax,Payez:2012,Wouters:2013,Ayala:2014,Tiwari:2017}. Only on small distance scales of order 100 Mpc, it has been speculated that supercluster magnetic field may directly affect the alignment of galaxies \citep{Tiwari:2016}. However, in this case also the alignment effect will depend on the level of randomness introduced by local effects within individual galaxies. 

\begin{figure}
	\includegraphics[width=\wx\textwidth]{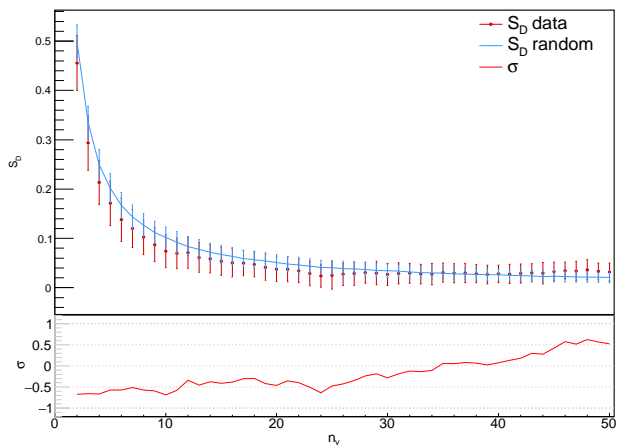}%
        \caption{The statistics $S_D$ calculated with 2D sample. Other details same as in Fig. \ref{fig:SD}}.
        \label{fig:SD_2D}
\end{figure}

The alignment of radio polarization reported earlier were usually in 2D due to unavailability of redshift, 
so to test if the alignment signal was a consequence of 2D projection, we calculate $S_D$ for our sample 
at fixed redshift, i.e. projecting all sources at same redshift. Even with 2D statistics we find the samples 
(all sources, quasar only sample)  agreeing well with random distribution (Fig. \ref{fig:SD_2D}).

\begin{figure}
	\includegraphics[width=\wx\textwidth]{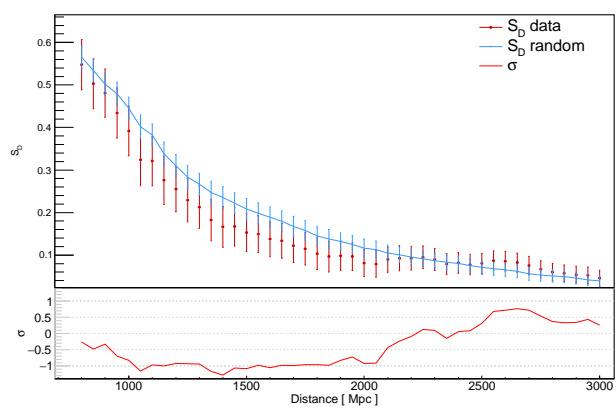}%
        \caption{Alignment at fixed distance. $S_D$ is average dispersion over the spheres of 
	radius `Distance' on x-axis at each source site. The number of nearest neighbour $\rm n_v$ 
	at each site varies here slightly. Other details same as in Fig. \ref{fig:SD}}.
        \label{fig:SD_dist}
\end{figure}

We also test the data with alternate statistics. In particularly we test the data by averaging $S_D$ at fixed 
distance i.e. averaging dispersion over the spheres of fixed radii at each sites. This also closely matches with 
random isotropic polarization distribution, results are shown in Fig. \ref{fig:SD_dist}. Furthermore, the high 
redshift quasar distribution is also isotropic. 
 

We have tested the large-scale radio polarization alignment signal with a robust and simultaneous dual frequency radio 
polarimetric survey \citep{Agudo:2014}. We do not find any observation of alignment at large-scale and polarization 
angles of AGNs and jet angles are fairly uniform and isotropic at scale equal and above $\sim$Gpc. 
However, due to low number density we are unable to probe scales less than 719 Mpc and 
cannot examine the alignment claimed at relatively smaller scales \citep{Tiwari:2013pol,Jagannathan:2014}. Furthermore 
with this sample we cannot explore the alignment with the axis of large quasar groups (LQGs) \citep{Pelgrims:2016}. 
This is because the LQGs are  typically of smaller size in comparison to distance scale of the first neighbour in our sample. 

We note that the jet position angle with respect to polarization angle in sample are not preferably parallel 
or perpendicular \citep{Agudo:2014} and this is somewhat unusual and unlike the predictions from 
axially symmetric jet models \citep{Falle:1991,Marti:1997,Komissarov:1999,Fendt:2012}. 
This probably indicate that the magnetic field or the particles in 
radio emission region are non axisymmetric and somehow due to local magnetic fields or other effects 
the polarization vector is misaligned  with respect to jet position angle. However, this observation does not change our 
conclusion as we find isotropy with jet position angle as well.

Nevertheless, this work adds to  the previous studies of large-scale radio polarization alignment anomalies. With this data 
we clearly see the isotropy above Gpc. These results further support the isotropy assumption in cosmology 
along with CMB and other supporting large-scale isotropy observations. 

\section{Acknowledgments}
This work is supported by NSFC Grants 1171001024 and 11673025,  the National Key Basic Research 
and Development Program of China (No. 2018YFA0404503) and the Science and Engineering Research Board (SERB), Government of India. The work is also supported by
NAOC youth talent fund 110000JJ01.
\bibliographystyle{aa}
\bibliography{master}
\end{document}